\documentclass[aps,prl,twocolumn]{revtex4-1}
\usepackage{epsfig}
\begin{document}
\author{Ming-Min Wang$^{1}$, Yu Jiang$^{1}$, Bin Wang$^{1}$, Wei-Min Sun$^{1,2}$,
and Hong-Shi Zong$^{1,2}$}
\address{$^{1}$ Department of Physics, Nanjing University,
Nanjing 210093, China}
\address{$^{2}$ Joint Center for Particle, Nuclear Physics
and Cosmology, Nanjing 210093, China}

\title{Calculation of bulk viscosity of QCD at zero temperature and
finite chemical potential}

\begin{abstract}
In this letter, based on Kubo's formula and the QCD low energy theorem,
we propose a direct formula for calculating the bulk viscosity of QCD at finite chemical potential $\mu$ and zero temperature. According to this formula, the bulk viscosity at finite $\mu$ is totally determined by the dressed quark propagator at finite $\mu$. We then use a dynamical, confining Dyson-Schwinger equation model of QCD to calculate the bulk viscosity at finite $\mu$. It is found that no sharp peak behavior of the bulk viscosity at finite $\mu$ is observed, which is quite different from that of the bulk viscosity at finite temperature.


\bigskip
\noindent
E-mail: zonghs@chenwang.nju.edu.cn.


\end{abstract}

\maketitle

Recently transport coefficients of hot/dense quark matter attract
lots of attentions. It is observed that the quark gluon plasma (QGP)
created at RHIC (Relativistic Heavy Ion Collider) behaves like a
nearly perfect fluid \cite{Arsene,Back,Adams,Adcox} with the ratio of the shear
viscosity $\eta$ over entropy density $s$ approaching the lower
bound of $1/4\pi$ \cite{sv1} near the phase transition temperature
$T_c$. Such a behavior indicates that the created QGP is not a weakly
coupled gas which is expected from perturbative quantum chromodynamics
(pQCD), but a strongly coupled liquid. On the other hand, lattice QCD
calculation shows that the ratio of the bulk viscosity $\zeta$ over
entropy density $s$ has a sharp peak near the critical temperature
\cite{bulk1,bulk2}. Such a sharp peak behavior of $\zeta$ has also
been observed in many model calculations including linear sigma
model \cite{bulk3}, real scalar model \cite{bulk4}, massless pion
gas model \cite{bulk5}, quasiparticle model \cite{bulk9}, $Z(2)$ and $O(4)$ models \cite{bulk6}, and Dyson-Schwinger equations \cite{jiang}. The bulk viscosity $\zeta$ characterizes the response of the system to the conformal transformation and vanishes when the system has a conformal equation of state, therefore the sharp peak of the bulk viscosity would strongly affect the physics of the QCD matter near $T_c$ and is very important for the study of QCD phase transition. 

Because the phase structure of QCD at finite density is much richer than that at finite temperature (see, for example, Ref. \cite{Cs1}), one may naturally ask whether the sharp peak behavior of $\zeta$ exists at zero temperature and finite $\mu$. As far as the present authors know, up to now there are only limited number of papers trying to answer the above question. For example, in Ref. \cite{bulk7} the authors study the viscosity at
finite $\mu$ with Nambu-Jona-Lasinio (NJL) model and in Ref.
\cite{bulk8} the authors study the viscosity of strange quark matter
at finite $\mu$ with quasi particle model. Just as will be shown below, elucidating the bulk viscosity of QCD matter requires a knowledge of the equation of state (EOS) of strongly interacting matter, and the plausibility of the results of any given study rest on the EOS employed, which is difficult to judge a priori (Presently, it is not feasible to study the EOS at finite $\mu$ rigorously in lattice simulation of QCD, due to the problems arising from the complex valued Euclidean fermion determinant at nonzero chemical potential). Consequently, the exploration of alternative equations of state and the identification of qualitatively consistent results is important. Here we would like to stress that the absence of confinement is a defect common to most existing model studies for the bulk viscosity of QCD matter \cite{bulk3,bulk4,bulk5,bulk9,bulk6,bulk7,bulk8}. A covariant, continuum approach incorporating both confinement and chiral symmetry in that direction is necessary. 

It is well known that the Dyson-Schwinger equations (DSEs) provide a nonperturbative, Poincare invariant continuum framework for analyzing QCD, i.e., for the study of confinement and dynamical chiral symmetry breaking (DCSB), and hadron observables \cite{DSE1,DSE2,DSE3,Alkofer,Fischer}. Herein we will employ a dynamical, confining Dyson-Schwinger equation model of QCD \cite{DSE10} to calculate the bulk viscosity of QCD at finite chemical potential and zero temperature. 

In order to make this letter self-contained, let us first recall the general approach of calculating bulk viscosity in existing literature. According to Kubo's formula the bulk viscosity $\zeta$ can be
expressed as \cite{bulk1}
\begin{eqnarray}
\zeta=-\frac{1}{9}\lim\limits_{\omega\rightarrow0}\frac{1}
{\omega}\mbox{Im}G^R(\omega,\vec{0})\label{bulkvis1},
\end{eqnarray}
where the retarded Green's function $G^R$ is defined as:
\begin{eqnarray}
G^R(x)&=&\int\frac{d\omega d^3\vec{k}}{(2\pi)^4}
G^R(\omega,\vec{k})e^{-i(\omega t
-\vec{k}\cdot\vec{x})}\nonumber\\
&=&-i\theta(t)\langle[\mathcal{T}_\mu^\mu(x),\mathcal{T}
_\mu^\mu(0)]\rangle
\end{eqnarray}
with $\mathcal{T}_{\mu\nu}$ the stress tensor of the system. In the
small frequency region one can assume the spectral density
$\rho(\omega,\vec{p})$ which is defined as
\begin{eqnarray}
\rho(\omega,\vec{p})=-\frac{1}{\pi}\mbox{Im}
G^R(\omega,\vec{p})\label{spectral1}
\end{eqnarray}
has the following form
\begin{eqnarray}
\frac{\rho(\omega,\vec{0})}{\omega}=\frac{9\zeta}{\pi}
\frac{\omega_0^2}{\omega_0^2+\omega^2},
\end{eqnarray}
where the parameter $\omega_0$ identifies the scale at which the
perturbation theory becomes valid \cite{bulk1}. With such ansatz the
bulk viscosity can be expressed as
\begin{eqnarray}
9\zeta\omega_0=2\int \limits_0^\infty
\frac{\rho(u,\vec{0})}{u}\,du
=\int d^4x\langle\,T_t\{\mathcal{T}_\mu^\mu(x)
\mathcal{T}_\mu^\mu(0)\}\rangle.
\end{eqnarray}

For QCD, the trace of stress tensor reads
\begin{eqnarray}
\mathcal{T}^\mu_\mu&=&m_q\bar{q}q+\frac{\beta(g)}{2g}
F_{\mu\nu}^aF^{a\mu\nu}
\equiv\mathcal{T}_F+\mathcal{T}_G\label{trace anomaly},
\end{eqnarray}
where $g$ is the strong coupling constant, $\mathcal{T}_F$ and $\mathcal{T}_G$ are the contribution of quark fields and of gluon field, respectively, and $\beta(g)$ is the QCD
$\beta$-function which determines the running behavior of $g$. In
Eq. (\ref{trace anomaly}) $q$ are quark fields with two flavors (in
this letter we will limit ourselves in two flavor case and set the
current quark mass $m_u=m_d=m$). From the QCD low-energy theorems at
finite temperature $T$ and $\mu$ \cite{LET1}, one can find
\begin{eqnarray}
\left(T\frac{\partial}{\partial
T}+\mu\frac{\partial}{\partial\mu}-d\right)\langle\hat{\mathcal
{O}}\rangle_T=\int d^4x\langle T_t\{\mathcal{T}_G(x)\hat\mathcal{O}(0)\}\rangle,
\end{eqnarray}
where $d$ is the canonical dimension of the operator
$\hat\mathcal{O}$. Using the above equation, one has
\begin{eqnarray}
\left(T\frac{\partial}{\partial
T}+\mu\frac{\partial}{\partial\mu}-4\right)\langle
\mathcal{T}_G\rangle_T 
=\int d^4x\langle T_t
\{\mathcal{T}_G(x)\mathcal{T}_G(0)\}\rangle,
\end{eqnarray}
\begin{eqnarray}
\left(T\frac{\partial}{\partial
T}+\mu\frac{\partial}{\partial\mu}-3\right)\langle
\mathcal{T}_F\rangle_T 
=\int d^4x\langle T_t
\{\mathcal{T}_G(x)\mathcal{T}_F(0)\}\rangle.
\end{eqnarray}
From the above two relations one obtains
\begin{eqnarray}\label{bulkvis2}
&&9\zeta\omega_0=\int d^4x\langle\,T_t\{\mathcal{T}_\mu^\mu(x)
\mathcal{T}_\mu^\mu(0)\}\rangle\nonumber\\
&=&\left(T\frac{\partial}{\partial
T}+\mu\frac{\partial}{\partial\mu}-4\right)\langle
\mathcal{T}_G\rangle_T+2\left(T\frac{\partial}{\partial
T}+\mu\frac{\partial}{\partial\mu}-3\right)\langle
\mathcal{T}_F\rangle_T \nonumber \\
&& +\int d^4x\langle\,T_t\{\mathcal{T}_F(x)
\mathcal{T}_F(0)\}\rangle\nonumber\\
&\approx& \left(T\frac{\partial}{\partial
T}+\mu\frac{\partial}{\partial\mu}-4\right)
\langle\mathcal{T}_\mu^\mu\rangle_T+\left(T\frac{\partial}{\partial
T}+\mu\frac{\partial}{\partial\mu}
-2\right)\langle\mathcal{T}_F\rangle_T \nonumber\\
&=&\left(T\frac{\partial}{\partial
T}+\mu\frac{\partial}{\partial\mu}-4\right)
(\mathcal{E}-3\mathcal{P}) \nonumber \\
&& +\left( T\frac{\partial}{\partial
T}+\mu\frac{\partial}{\partial\mu}-2\right)
\langle\mathcal{T}_F\rangle_T,
\end{eqnarray}
where $\mathcal{E}$ is the energy density and $\mathcal{P}$ is the
pressure density of QCD. Here, because the current quark mass $m$ of u and d quark is very small, in deriving Eq. (\ref{bulkvis2}) we have neglected the term proportional to $m^2$. 

Now we take the $T\rightarrow0$ limit of the above expression. In this limit the terms involving $T\partial/\partial T$ vanish. Therefore one
would obtain the following expression of $\zeta$ at zero $T$ and finite $\mu$:
\begin{eqnarray}
9\zeta\omega_0=-4\mu\rho_q(\mu)+16\mathcal{P}(\mu)+2m\mu\frac
{\partial\langle\bar{q}q\rangle}{\partial\mu}
-4m\langle\bar{q}q\rangle,\label{bulkvis3}
\end{eqnarray}
where $\rho_q(\mu)$ is the quark number density which can be written as
(see, for example, Ref. \cite{zong1})
\begin{eqnarray}
\rho_q(\mu)=\frac{\partial\mathcal
{P}(\mu)}{\partial\mu}\label{qns1}
=-N_cN_f\int\frac{d^4p}{(2\pi)^4}\mbox{tr}
\left\{G_q[\mu](p)\gamma_4\right\}\label{qns2}
\end{eqnarray}
and $\langle\bar{q}q\rangle$ is the quark condensate at zero $T$ and finite $\mu$ which can be defined as
\begin{equation}\label{quarkcon}
\langle\bar{q}q\rangle=-N _c N_f \int \frac{d^4 q}{(2\pi)^4}\mathrm{tr}\{ G_q[\mu](p)\}.
\end{equation}
In Eqs. (\ref{qns2}) and (\ref{quarkcon}) $G_q[\mu](p)$ is the dressed quark propagator at finite $\mu$, $N_c$ and $N_f$ denote the number of colors and of flavors, respectively, and the trace operation tr is over Dirac indices (we will not write the renormalization constants explicitly because one would find in the final result the renormalization constants cancel each other). 

Eq. (\ref{bulkvis3}) indicates that the bulk viscosity of QCD at zero $T$ and finite $\mu$ is determined by $\rho_q(\mu)$, the pressure density $\mathcal{P}(\mu)$ and the quark condensate at finite $\mu$. From Eqs. (\ref{qns2}) and (\ref{quarkcon}) it can be seen that the quark number density and quark condensate at finite $\mu$ is totally determined by the dressed quark propagator at finite $\mu$. As will be shown below, the pressure density is also totally determined by the dressed quark propagator at finite $\mu$. This means that the bulk viscosity at finite $\mu$ is totally determined by the dressed quark propagator at finite $\mu$. This is quite different from the
case of finite $T$ in which the term involving the speed of sound
contributes to the peak behavior of $\zeta/s$ around $T_c$
\cite{bulk1}.

From Ref. \cite{zong1} one can obtain the following:   
\begin{eqnarray}
\mathcal{P}(\mu)&=&\mathcal{P}(\mu=0)+
\int_0^\mu\,\rho_q(\mu^\prime)d\mu^\prime=\mathcal{P}(\mu=0)\nonumber\\
&-&N_cN_f\int_0^\mu d\mu^\prime \int\frac{d^4p}{(2\pi)^4}\mbox{tr}
\left\{G_q[\mu^\prime](p)\gamma_4\right\}\label{pressure1}.
\end{eqnarray}
Based on the above equation it can be seen that the pressure density
$\mathcal{P}(\mu)$ is the sum of two terms: the first term
$\mathcal{P}(\mu=0)$ (the pressure density of vacuum) is only a
$\mu$-independent term; the second term, which is totally determined
by $G_q[\mu](p)$, contains all the nontrivial $\mu$-dependence. Just as was shown in Ref. \cite{zong1}, the constant term $\mathcal{P}(\mu=0)$ can be calculated through Cornwall-Jackiw-Tomboulis effective
action \cite{CJT}. However, in the present paper we are only interested in the $\mu$-dependence of $\zeta$, so we ignore this term. 

According to Eqs. (8-10) and Eq. (\ref{pressure1}), once the dressed quark propagator at finite $\mu$ is known, one can obtain
the bulk viscosity of QCD at finite $\mu$. However, at present it is very difficult to calculate $G_q[\mu](p)$ from first principles of QCD
and one has to resort to various QCD models. Over the past few
years, considerable progress has been made in the framework of the
rainbow-ladder approximation of the DSEs approach
\cite{DSE1,DSE2,DSE3,Alkofer,Fischer}, which provides a successful description
of various nonperturbative aspects of strong interaction physics. In
the rest of this paper we will use DSEs approach to calculate
$\zeta$.

To obtain the dressed quark propagator at finite $\mu$ let us first
recall the following general result given in Refs.
\cite{zong2,Hou,zong3,Feng1,Feng2}: If one adopts the rainbow approximation of the
DSEs (rainbow approximation means one uses the bare vertex to approximate the full 
quark-gluon vertex in the quark DSE)
and ignores the $\mu$ dependence of
the dressed gluon propagator (one often makes this approximation in the study of finite density QCD because it is usually believed that the effect of
chemical potential on the gluon propagator arising from quark-loop
insertions is small in comparison with that on the quark propagator;
this is a commonly used approximation
in studying the dressed quark propagator at finite $\mu$
\cite{DSE2,zong2,Hou,zong3,Feng1,Feng2,DSE5,DSE6,DSE7,DSE8,DSE9}), then one can
obtain the dressed quark propagator at finite $\mu$ from the one at
$\mu=0$ by the following substitution (more details can be found in
Refs. \cite{zong2,Hou,zong3,Feng1,Feng2}):
\begin{equation}\label{propfinitemu}
G_q[\mu](p)=G_q[\mu=0]({\tilde p}),\label{prp1}
\end{equation}
where $\tilde{p}=(\vec{p},p_4+i\mu)$. 
The validity of this approximation has been discussed in detail in Refs. \cite{zong3,Feng1,Feng2,JiangShi} and has been applied to the study of properties of quark stars \cite{Lihua,Li1}. 
Therefore, once the dressed quark propagator at $\mu=0$ is known, the dressed quark propagator
at finite $\mu$ and $T=0$ can be obtained by means of Eq. (\ref{prp1}).

Now one needs to specify the form of the dressed quark propagator at
$\mu=T=0$. In Ref. \cite{DSE10}, guided by the solution of the
coupled set of DSEs for the ghost, gluon and quark propagator in the
Landau gauge, the following meromorphic form of the dressed quark
propagator is proposed:
\begin{equation}
\label{prp2} G_q(p)=\sum_{j=1}^{n_P}\left(\frac{r_j}
{i{\not\!p}+a_j+ib_j}+\frac{r_j}{i{\not \!p}+a_j-ib_j}\right).
\end{equation}
The propagator of this form has $n_P$ pairs of complex conjugate
poles located at $a_j\pm ib_j$. When some $b_j$ is set to zero, the
pair of complex conjugate poles degenerates to a real pole. The
residues $r_j$ are real (note that a similar meromorphic form of the quark propagator was previously proposed in Ref. \cite{Tandy}, in which the residues in the two additive terms are complex conjugate of each other). The values of parameters in Eq. (\ref{prp2}) are listed in Table I \cite{DSE10} where 2CC means two pairs of complex conjugate poles, 1R1CC means one real pole and one pair of complex conjugate poles and 3R means three real poles.
The parameters of these meromorphic forms in Table I are fitted by requiring that these meromorphic forms of propagator reproduce some physical observables (the two-quark condensate
and pion decay constant) and lattice data (the zero momentum values of the mass function and wave-function renormalization, $M_0$ and $Z_0^f$, and an approximate width of the region of large dynamical mass generation, $\omega_L$) (for more details, please see \cite{DSE10}). 
Here we should stress that the real
dressed quark propagator has much more complicated analytic
structures than that given in Eq. (\ref{prp2}). However, despite its simplicity, this kind of
parametrization of the dressed quark propagator has several appealing features: it can simultaneously describe DCSB and quark confinement which are very important for low energy QCD; at large momenta it reduces to the free fermion propagator, as required by asymptotic freedom. Phenomenologically,  
these parameterizations of the dressed quark propagator can give a quite good
description of low energy QCD \cite{DSE10}.

\begin{widetext}
\begin{center}
{\scriptsize Table I. The parameters for the quark propagator. These
parameters are taken directly from Table II of Ref. \cite{DSE10}}.
\scriptsize \vspace{0.1cm}

\begin{tabular*}{16cm}{l@{\extracolsep{\fill}}*{8}{c}} \hline\hline
Parameterization&$r_1$&$a_1$ (GeV)&$b_1$ (GeV)&$r_2$&$a_2$
(GeV)&$b_2$ (GeV)&$r_3$&$a_3$
(GeV)\\
\hline 2CC&0.360&0.351&0.08&0.140&-0.899&0.463&-&-\\
\hline 1R1CC&0.354&0.377&-&0.146&-0.91&0.45&-&-\\
\hline 3R&0.365&0.341&-&1.2&-1.31&-&-1.06&-1.40\\
\hline\hline
\end{tabular*}
\end{center}
\end{widetext}

With such propagator the quark number density and the
$\mu$-dependent term of $\mathcal{P}(\mu)$ in Eq. (\ref{pressure1})
can be calculated as \cite{zong1}
\begin{eqnarray}
\rho_q(\mu)&=&\frac{2N_c
N_f}{3\pi^2}\sum_{j=1}^{n_P}r_j\theta(\mu-|a_j|)\nonumber\\
&&\times\left(\mu^2-\frac{a_j^2b_j^2}
{\mu^2}-a_j^2+b_j^2\right)^{\frac{3}{2}} ,\label{qns3}\\
\int_0^\mu\,\rho_q(\mu^\prime)d\mu^\prime&=&\frac{2N_cN_f}{3\pi^2}
\sum_{j=1}^{n_P}r_j\theta(\mu-|a_j|)I_j\label{pressure2},
\end{eqnarray}
where $I_j$ reads:
\begin{eqnarray}
I_j&\equiv&\frac{3(a_j^4+b_j^4-6a_j^2b_j^2)}
{16}\ln\frac{\sqrt{\mu^2+b_j^2}+\sqrt{\mu^2-a_j^2}}
{\sqrt{\mu^2+b_j^2}-\sqrt{\mu^2-a_j^2}}\nonumber\\
&&+\frac{3(a_j^2-b_j^2)|a_jb_j|}{2}\arctan\sqrt{
\frac{b_j^2(\mu^2-a_j^2)}{a_j^2(\mu^2+b_j^2)}}\nonumber\\
&&+\left[\frac{\mu^2}{4}-\frac{5}{8}(a_j^2-b_j^2)\right]
\sqrt{(\mu^2-a_j^2)(\mu^2+b_j^2)}\nonumber\\
&&+\frac{a_j^2b_j^2}{2\mu^2}\sqrt{(\mu^2-a_j^2)(\mu^2+b_j^2)}.
\end{eqnarray}
From Eq. (\ref{qns3}) it can be seen that when $\mu$ 
is below a critical value $\mu_c=\mathrm{min}\{|a_j|\}$ ($\mu_c=351~\mathrm{MeV}, 377~\mathrm{MeV}$ and $341~\mathrm{MeV}$ for the 2CC, 1R1CC and 3R parametrization, respectively), the quark number density vanishes identically. Namely, $\mu=\mu_c$ is a singularity which separates two regions with different quark number densities. This result agrees qualitatively with the general conclusion
of Ref. \cite{PHASE}. In that reference, based on a universal argument,it is pointed out that the existence of some singularity at the point $\mu=\mu_c$ and $T=0$ is a robust and model-independent prediction. The numerical value of the critical chemical potential in pure QCD (i.e., with electromagnetic interaction being switched off) is estimated to be $307~\mathrm{MeV}$. The value of $\mu_c$ obtained here is of the same order of magnitude as the estimates in Ref. \cite{PHASE}.

The quark condensate at finite $\mu$ with dressed quark propagator
given in Eq. (\ref{prp2}) has been obtained in Ref. \cite{CON}:
\begin{eqnarray}
\label{Con1}\langle \bar{q}q\rangle=\langle
\bar{q}q\rangle_0+\frac{2N_cN_f}{\pi^2}\sum_{j=1}^{n_P}
r_ja_j\theta(\mu-|a_j|)f_j(\mu),
\end{eqnarray}
where $\langle \bar{q}q\rangle_0$ is the quark condensate at $\mu=0$
and $f_j$ reads
\begin{eqnarray}
f_j&\equiv&\frac{1}{2}\left(1-\frac{b_j^2}{\mu^2}\right)
\sqrt{(\mu^2-a_j^2)(\mu^2+b_j^2)}\nonumber\\
&&+\frac{3b_j^2-a_j^2}{4}
\ln\frac{\sqrt{\mu^2+b_j^2}+\sqrt{\mu^2-a_j^2}}
{\sqrt{\mu^2+b_j^2}-\sqrt{\mu^2-a_j^2}}\nonumber\\
&&+\frac{b_j^2-3a_j^2}{2}\left|\frac{b_j}{a_j}\right|\arctan
\sqrt{\frac{b_j^2(\mu^2-a_j^2)}{a_j^2(\mu^2+b_j^2)}}.
\end{eqnarray}\\
From Eq. (\ref{Con1}) it can also be seen that when $\mu<\mu_c$, the quark condensate equals its vacuum value.  

Substituting Eqs. (\ref{qns3}), (\ref{pressure2}) and (\ref{Con1})
into Eq. (\ref{bulkvis3}) one can calculate the bulk viscosity at
zero $T$ and finite $\mu$. The numerical result is shown in Fig. 1.
\begin{figure}
\includegraphics[width=6cm]{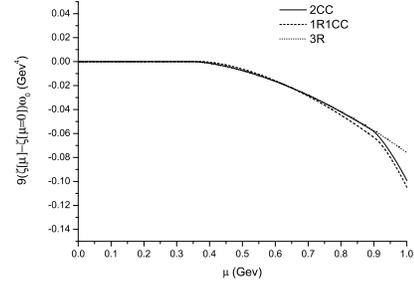}\\
\vspace{-0.6cm}
\caption{Bulk viscosity at finite $\mu$}\label{figBV1}
\end{figure}
From Fig. 1 one can find the bulk viscosity is kept unchanged from the
value at $\mu=0$ when $\mu$ is smaller than 
$\mu_c$, and $\mu_c$ is the point when quarks
get populated (like the liquid-gas point, if there were a self-bound nuclear or 
constituent-quark groundstate). This is understandable because as was analyzed in Ref.
\cite{zong1}, when $\mu<\mu_c$, the ground state at finite $\mu$ remains the vacuum and the physical quantities should be kept unchanged. When $\mu>\mu_c$, it can be seen from Fig. 1 that the bulk viscosity decreases monotonously as $\mu$ increases and no sharp peak is observed at finite $\mu$. This behavior of bulk viscosity at finite $\mu$ does not show a first-order chiral phase transition (or a rapid cross-over), which is quite different from that at finite temperature. The result of such a behavior is qualitatively consistent
with the result obtained in Ref. \cite{bulk7}. Here, we would like to point out that the NJL model \cite{Klevansky, Buballa} is the most widely adopted nonperturbative QCD model in the study of finite density QCD at present. The NJL model can phenomenologically incorporate DCSB but does not have quark confinement, whereas the DSEs approach can simultaneously incorporate DCSB and quark confinement. 
Over the past few years considerable progress has been made in the framework of
the rainbow-ladder approximation of the DSEs approach at finite temperature and finite density
(see, e.g., the review article Ref. \cite{DSE2}). Due to this we naturally expect that the DSEs approach can be applied to the study of bulk viscosity of QCD. This is our main motivation to use the DSEs approach to calculate the bulk viscosity of QCD at finite density in this work.
Finally, it should be noted that that in obtaining the result of bulk viscosity in this paper, we have made use of 
Eq. (\ref{propfinitemu}). However, as was pointed out in 
Refs. \cite{zong2,Hou,zong3,Feng1,Feng2}, Eq. (\ref{propfinitemu}) is obtained under the approximation that 
one neglects the $\mu$ dependence of the dressed gluon propagator. This approximation is valid only in small $\mu$ region. Therefore, if one applies Eq. (\ref{propfinitemu}) to the case of large $\mu$, one should be cautious. 

To summarize, in this letter, based on the Kubo's formula and the QCD low energy theorem, we have derived a direct formula for calculating the bulk viscosity of QCD at finite chemical potential $\mu$ and zero temperature. According to this formula, the bulk viscosity at finite 
$\mu$ is totally determined by the dressed quark propagator at finite 
$\mu$. Then, under the rainbow approximation of DSEs and ignoring the $\mu$-dependence of the dressed gluon propagator, we use the meromorphic quark propagator proposed in Ref. \cite{DSE10} to calculate the bulk viscosity at finite $\mu$. It is found that when $\mu$ is below a critical value $\mu_c$, the bulk viscosity equals its vacuum value. When $\mu>\mu_c$, the bulk viscosity decreases monotonously as $\mu$ increases, and no sharp peak is observed at finite $\mu$. 
This means that in our model calculation in the framework of DESs approach the bulk viscosity at finite $\mu$ does not show a first-order chiral phase transition (or a rapid cross-over), which is quite different from that at finite temperature. 

\begin{acknowledgments}
This work is supported in part by the National Natural Science Foundation of China (under Grant Nos. 10775069, 10935001 and 11075075) and the Research Fund for the Doctoral Program of Higher Education (Grant No. 200802840009).
\end{acknowledgments}


\begin{thebibliography}{99}
\bibitem{Arsene} I. Arsene, \emph{et. al.}, Nucl. Phys. \textbf{A 757}, 1 (2005).
\bibitem{Back} B. B. Back, \emph{et. al.}, Nucl. Phys. \textbf{A 757}, 28 (2005).
\bibitem{Adams} J. Adams, \emph{et. al.}, Nucl. Phys. \textbf{A 757}, 102 (2005).
\bibitem{Adcox} K. Adcox, \emph{et. al.}, Nucl. Phys. \textbf{A 757}, 184 (2005).
\bibitem{sv1} G. Policastro, D. T. Son, A. O. Starinets, Phys. Rev.
Lett. \textbf{87}, 081601 (2001).
\bibitem{bulk1} F. Karsch, D. Kharzeev, and K. Tuchin, Phys. Lett. B
\textbf{663}, 217 (2008).
\bibitem{bulk2} H. B. Meyer, Phys. Rev. Lett. \textbf{100}, 162001
(2008).
\bibitem{bulk3} K. Paech, and S. Pratt, Phys. Rev. \textbf{C 74},
014901 (2006).
\bibitem{bulk4} B. C. Li, and M. Huang, Phys. Rev. \textbf{D 78},
117503 (2008).
\bibitem{bulk5} J. W. Chen, and J. Wang, Phys. Rev. \textbf{C 79},
044913 (2009).
\bibitem{bulk9} C. Sasaki, and K. Redlich, Phys. Rev. \textbf{C 79}, 055207 (2009).
\bibitem{bulk6} B. C. Li, and M. Huang, Phys. Rev. \textbf{D 80}, 034023 (2009).
\bibitem{jiang} Y. Jiang, B. Wang, W. M. Sun, and H. S. Zong, Mod. Phys. Lett. {\bf A25}, 1689 2010.
\bibitem{Cs1} M. G. Alford, A. Schmitt, K. Rajagopal, and T. Sch\"{a}fer, Rev. Mod. Phys. \textbf{80}, 1455 (2008).
\bibitem{bulk7} C. Sasaki, and K. Redlich, Nucl. Phys. \textbf{A 832}, 62 (2010).
\bibitem{bulk8} X. P. Zheng, M. Kang, X. W. Liu, and S. H. Yang,
Phys. Rev. \textbf{C 72}, 025809 (2005).
\bibitem{DSE1} C. D. Roberts and A. G. Williams, Prog. Part. Nucl.
Phys. \textbf{33}, 477 (1994), and references therein.
\bibitem{DSE2} C. D. Roberts and S. M. Schmidt, Prog. Part. Nucl.
Phys. \textbf{45S1}, 1 (2000), and references therein.
\bibitem{DSE3} P. Maris and C. D. Roberts, Int. J. Mod. Phys.
\textbf{E 12}, 297 (2003).
\bibitem{Alkofer} R. Alkofer and L. von Smekal, Phys. Rept.
\textbf{353}, 281 (2001), and references therein.
\bibitem{Fischer} C. S. Fischer and R. Alkofer, Phys. Rev. \textbf{D 67}, 094020 (2003), and references therein.
\bibitem{DSE10} R. Alkofer, W. Detmold, C.S. Fischer, and P. Maris, Phys. Rev. D
{\bf 70}, 014014 (2004).
\bibitem{LET1} I. A. Shushpanov, J. I. Kapusta, and P. J. Ellis,
Phys. Rev. C {\bf 59}, 2931 (1999).
\bibitem{zong1} H. S. Zong, and W. M. Sun, Phys. Rev. D {\bf 78},
054001 (2008).
\bibitem{CJT} J. M. Cornwall, R. Jackiw, and E. Tomboulis,
 Phys. Rev. D {\bf 10}, 2428 (1974).
\bibitem{zong2} H. S. Zong, L. Chang, F. Y. Hou, W. M. Sun and Y. X. Liu, Phys.
Rev. C {\bf 71}, 015205 (2005).
\bibitem{Hou} F. Y. Hou, L. Chang, W. M. Sun, H. S. Zong and Y. X. Liu, Phys. Rev. C{\bf 72}, 034901 (2005).
\bibitem{zong3} H. T. Feng, F. Y. Hou, X. He, W. M. Sun and H. S. Zong, Phys.
Rev. D {\bf 73}, 016004 (2006). 
\bibitem{Feng1}H.T. Feng, W.M. Sun, D.K. He, and
H. S. Zong, Phys. Lett. B {\bf 661}, 57 (2008).
\bibitem{Feng2} H. T. Feng, M. He, W. M. Sun, and H. S. Zong, Phys. Lett. {\bf B688}, 178 (2010).
\bibitem{DSE5} Y. Taniguchi and Y. Yoshida, Phys. Rev. D {\bf 55}, 2283 (1997).
\bibitem{DSE6} D. Blaschke, C. D. Roberts and S. Schmidt, Phys. Lett. B {\bf 425}, 232 (1998).
\bibitem{DSE7} P. Maris, C. D. Roberts and P. C. Tandy, Phys. Lett. B {\bf 420}, 267 (1998).
\bibitem{DSE8} A. Bender, W. Detmold and A. W. Thomas, Phys. Lett. B {\bf 516}, 54 (2001).
\bibitem{DSE9} O. Miyamura, S. Choe, Y. Liu, T. Takaishi, and A. Nakamura, Phys. Rev. D {\bf 66}, 077502 (2002).
\bibitem{JiangShi} Y. Jiang, Y.M. Shi, H.T. Feng, W.M. Sun, and H.S. Zong, Phys. Rev. C {\bf 78}, 025214 (2008).
\bibitem{Lihua} H. Li, X.L. Luo, and  H.S. Zong, Phys. Rev. D {\bf 82}, 065017 (2010).
\bibitem{Li1} H. Li, X. L. Luo, Y Jiang, and H. S. Zong, Phys. Rev. {\bf D 83}, 025012 (2011).
\bibitem{Tandy} M. S. Bhagwat, M. A. Pichowsky, and P. C. Tandy, Phys. Rev. D {\bf 67}, 054019 (2003).
\bibitem{PHASE} M.A. Halasz, A.D. Jackson, R.E. Shrock, M.A. Stephanov, and J.J.M. Verbaarschot, Phys. Rev. D {\bf 58}, 096007 (1998).
\bibitem{CON} Y. Jiang, Y. B. Zhang, W. M. Sun, and H. S. Zong,
Phys. Rev. D {\bf 78}, 014005 (2008).
\bibitem{Klevansky} S.P. Klevansky, Rev. Mod. Phys. {\bf 64}, 649 (1992), and references therein.
\bibitem{Buballa} M. Buballa, Phys. Rept. {\bf 407}, 205 (2005), and references therein. 




\end{thebibliography}
\end{document}